\documentclass[11pt,letterpaper]{article}
\pdfoutput=1
\usepackage{amscd}
\usepackage{amsmath, bbm} 

\usepackage{jheppub} 
\usepackage{slashed}
\usepackage{multirow}
\usepackage{array}
\usepackage{graphicx}
\usepackage{tikz-feynman}
\usepackage{tikz-cd}
\usepackage{longtable}
\usepackage{setspace}

\usepackage[utf8]{inputenc}
\usepackage[english]{babel}

\newtheorem{proposition}{Proposition}[section]
\newtheorem{conjecture}{Conjecture}[section]
\newtheorem{corollary}{Corollary}[proposition]

\allowdisplaybreaks
\newcommand\mybox[1]{\parbox[t]{0.9\textwidth}{%
    \setstretch{1.33}\raggedright$\displaystyle #1$}}

\title{Lorentz and permutation invariants of particles II} 

\author[a]{Ben Gripaios,}
\author[b]{Ward Haddadin,}
\author[a]{and C. G. Lester}
\affiliation[a]{Cavendish Laboratory, University of Cambridge, J.J. Thomson Avenue, Cambridge, CB3 0HE, United Kingdom} 
\affiliation[b]{DAMTP, University of Cambridge, Wilberforce Road, Cambridge, CB3 0WA, United Kingdom}

\emailAdd{gripaios@hep.phy.cam.ac.uk}
\emailAdd{w.haddadin@damtp.cam.ac.uk}
\emailAdd{lester@hep.phy.cam.ac.uk}
\abstract{Two theorems of Weyl tell us that the algebra of Lorentz- (and parity-)
  invariant polynomials in the momenta of $n$ particles are generated
  by the dot products and that the redundancies which arise when $n$ exceeds the spacetime dimension $d$ are generated by the $(d+1)$-minors of the $n \times n$ matrix of dot products. Here, we use the Cohen-Macaulay structure of the invariant algebra to provide a more direct characterisation in terms of a Hironaka decomposition. Among the benefits of this approach is that it can be generalized straightforwardly to cases where a permutation group acts on the particles, such as when some of the particles are identical. In the first non-trivial case, $n=d+1$, we give a homogeneous system of parameters that is valid for the action of an arbitrary permutation symmetry and make a conjecture for the full Hironaka decomposition in the case without permutation symmetry. An appendix gives formul\ae\ for the computation of the relevant Hilbert series for $d \leq 4$.
}

\newcommand{\R}{\mathbb{R}}
\newcommand{\C}{\mathbb{C}}
\newcommand{\Z}{\mathbb{Z}}

\begin{document} 
\maketitle
\flushbottom

\section{Introduction}\label{sec:Intro}
Consider the momenta $p_i, i \in \{1,2,\dots,n\}$, of $n$ particles living in $d$ spacetime dimensions.
By allowing the momenta to take values in the complex numbers rather than the reals, we can replace the action of the Lorentz group including parity transformations with the action of the orthogonal group $O(d,\C)$. Weyl's First Fundamental Theorem (FFT) \cite{Weyl} states that the algebra of invariant polynomials in the $p_i$'s is generated by 
the $n(n+1)/2$ dot products $p_i \cdot p_j$.\footnote{Without parity transformations, the group becomes $SO(d, \C)$, and we have additional generators given by the possible contractions of the epsilon tensor with the momenta. We postpone the discussion of this case to future work \cite{LPIPIII}.} Weyl's Second Fundamental Theorem (SFT) characterises the relations between the generators: when $n \leq d$ there are no relations (so the dot products are algebraically independent and the algebra of invariants is a polynomial algebra), while when $n > d$, the relations are generated by the $(d+1)$-minors\footnote{We define a $(d+1)$-minor of a matrix to be the determinant of a $(d+1) \times (d+1)$ submatrix.} of the $n\times n$ matrix whose entries are given by $p_i \cdot p_j$.

In previous work \cite{LPIPI},\footnote{Similar ideas were explored in \cite{Melia}, in the context of classifying higher-dimensional operators in effective scalar field theories.} we generalized the FFT to include the action of an arbitrary group of permutations of the $n$ particles. (This is relevant, for example, when some of the particles are indistinguishable, which is an inevitable consequence of quantum field theory.) 
A major difference is that, even when $n \leq d$,  the algebra of invariants is not a polynomial algebra once we include permutations. This simple observation already suggests that attempts to generalise the SFT to the case where permutations are included will lead to unpleasantness.

In this work, we replace the FFT and SFT by a more direct description of the algebra of Lorentz- and permutation-invariants, using tools of commutative algebra which were not available to Weyl. In particular, we use the fact that (via a theorem of Hochster and Roberts \cite{Hochster}) the algebra of invariants is Cohen-Macaulay, and so admits a Hironaka decomposition as a free, finitely-generated module over a polynomial subalgebra.\footnote{For the necessary definitions, we refer the reader to \cite{LPIPI}.} Thus, a direct description of the invariant algebra can be given in terms of a set of generators of such a polynomial subalgebra, termed either {\em primaries} or a {\em homogeneous system of parameters} (HSOP), and a set of basis elements for the module, called  {\em secondaries}. In particular, every element in the algebra can be expressed {\em uniquely} in terms of primaries and secondaries, and multiplication in the algebra is completely encoded in the finite set of products of secondaries.

Expressing the invariant algebra in terms of a Hironaka decomposition in this way has multiple potential benefits over using any old set of generators. As an example, consider the problem where one wishes to assign the task of analysing experimental data to an unintelligent computer using these generators via some (machine learning) algorithm. If done through any old set of generators, the algorithm could waste considerable effort trying to differentiate between two quantities, before learning that they do, in fact, coincide. By using a Hironaka decomposition of the algebra, where every element is uniquely expressed, this waste is automatically avoided.

As in the previous work, the difficulty is in finding these Hironaka decompositions explicitly. In the Sections that follow, we proceed to sketch out the background results of invariant theory and employ them to find Hironaka decompositions in the first non-trivial case, {\em viz.} $n = d+1$. We solve the hardest step in the procedure, namely to find HSOPs. We do this both for the case without permutations and for the case with all permutations included; the latter serves as a HSOP for an arbitrary subgroup of permutations. Unfortunately, even though the remaining step of finding the secondaries reduces to a conceptually straightforward exercise in linear algebra, the available algorithm proceeds by brute-force Groebner basis methods \cite{dk} and runs out of steam in cases with more than a few particles. But we hope that our results, modest though they are, will inspire others to make more targeted attacks on the problem.  In Section \ref{sec:Examples}, we present Hironaka decompositions of the cases with  $(n,d) = (5,4)$ with no permutations included, $(n,d) = (3,2)$ with all permutations included, and a conjecture for the Hironaka decomposition of the general case of $n = d + 1$ with no permutations. An Appendix gives the details of the relevant Hilbert series computations.

\subsection{Technical statement of results} \label{sec:Technical}

Let us now give a more technical statement of the results. It is convenient, for a variety of reasons, to regard the momenta as taking values in a
vector space $V \cong \C^{nd}$ over the algebraically-closed field of
complex numbers. Doing so not only leads to simplifications on the
commutative algebra side, but also allows us to replace the Lorentz
group by its complexification $O(d, \C)$. The polynomials in the momenta
then form an algebra over $\C$, which we denote $\C[V]$ and the
Lorentz- and permutation-invariant polynomials, for some permutation group $P \subseteq S_n$, form a subalgebra $\C[V]^{O(d) \times P}
\subset \C[V]$. Phrased in these terms, the FFT is the statement that, in the case where $P$ is the trivial group, there exists a surjective algebra map $W:\C[y_{ij}] \twoheadrightarrow \C[V]^{O(d)}$, where $\C[y_{ij}]$ is the polynomial algebra over $\C$ in the variables $y_{ij}, i,j \in
\{1,\dots,n\}, i\leq j$, given explicitly on the generators by
$W: y_{ij} \mapsto p_i \cdot p_j$ and extended to an arbitrary polynomial in the obvious way. The SFT is then the statement that the kernel of this map, $\ker W$, is non-trivial when $n > d$, being the ideal $I \subset \C[y_{ij}]$ generated by the $(d+1)$-minors of the matrix whose $ij$-th entry is $y_{ij}$ for $i \leq j$ and $y_{ji}$ for $i > j$.

Now we consider the action of an arbitrary permutation group $P \subseteq S_n$ on the momenta $p_i$ and the variables $y_{ij}$ in the usual way (where again we make the replacement $y_{ij} = y_{ji}$ when $i> j$). One can show that the restricted Weyl map from the permutation-invariant subalgebra $\C[y_{ij}]^P \subset \C[y_{ij}]$ surjects onto the Lorentz- and permutation-invariant subalgebra, $W|: \C[y_{ij}]^P \twoheadrightarrow \C[V]^{O(d) \times P}$. The obvious generalisation of the FFT for the orthogonal group to the one including permutations is to provide a set of generators of the algebra $\C[y_{ij}]^P$. In previous work \cite{LPIPI}, we constructed a general method for finding such a set. What our generalisation of Weyl's work did not include is the generalisation of the SFT to characterise the kernel of the restriction map $\ker W|$. Formally, the kernel is the intersection of the ideal of relations with the invariant algebra, $\ker W| = I \cap \C[y_{ij}]^P$. In practice however, it is difficult to explicitly describe $\ker W|$, for a couple of reasons. For one thing, as stated previously, whereas $\C[y_{ij}]$ is a polynomial algebra, the invariant algebra $\C[y_{ij}]^P$ has a more complicated structure in general: it is Cohen-Macaulay and therefore can be expressed as a free, finitely-generated algebra over a polynomial subalgebra. For another, it turns out that the generators of the ideal $I$ transform in an unpleasant representation of the permutation group, making finding the corresponding permutation-invariant generators difficult.

We therefore follow an alternative approach, seeking a more direct description of the Lorentz- and permutation-invariant algebra $\C[V]^{O(d) \times P}$. The {\em Hochster-Roberts} theorem \cite{Hochster} states that an invariant algebra $K[V]^G$ is Cohen-Macaulay if $G$ is a linearly reductive group.\footnote{A linear algebraic group $G$ is called {\em linearly reductive} if for every rational representation $V$ and every $v \in V^G \setminus \{0\}$, there exists a linear invariant function $f \in (V^*)^G$ such that $f(v) \neq 0$.} Here and elsewhere in this work, $K$ will denote a field of characteristic zero. Since $O(d, \C) \times P$ is linearly reductive, the theorem applies and the algebra $\C[V]^{O(d) \times P}$ can be expressed as a free, finitely-generated module over a polynomial subalgebra. That is, the algebra can be expressed in terms of a Hironaka decomposition as $\C[V]^{O(d) \times P} = \bigoplus_k \eta_k \C[\theta_l]$ where the $\{\eta_k\}$ are the secondaries, the $\{\theta_l\}$ form a HSOP, and multiplication in the algebra is uniquely defined via $\eta_k \eta_m = \sum_j f^j_{km} \eta_j$, with $f^j_{km} \in \C[\theta_l]$. Every element in the algebra is then uniquely expressed as a linear sum of secondaries with coefficients which are polynomials in the HSOP.

The difficult part of finding Hironaka decompositions begins when one tries to find valid HSOPs as, apart from using inefficient algorithms \cite{Buchberger}, there is no obvious way to obtain them. Furthermore, the properties that a valid HSOP needs to satisfy are non-trivial and difficult to check. In \cite{LPIPI}, we were able to sidestep this by repurposing Gauss's results on permutation-invariants. Here, we are not so lucky. In Section \ref{sec:HSOPS}, we propose HSOPs for the algebras $\C[V]^{O(d) \times P}$, with $n = d + 1$, in the two cases where $P = 1$ (with 1 denoting the trivial group) and $P=S_n$ and explicitly verify that they satisfy the necessary conditions.

\section{HSOPs for $\C[V]^{O(d) \times P}$} \label{sec:HSOPS}

In this Section, we find HSOPs for the algebras $\C[V]^{O(d) \times P}$ in the $n = d+1$ case with no permutation symmetry, $P=1$, and with full permutation symmetry, $P = S_n$. In fact, the HSOP for $P = S_n$ is also a HSOP for any $P \subset S_n$, and so we obtain a complete solution of this part of the problem.

The necessary conditions for a set of polynomials to constitute a HSOP are twofold: firstly, the polynomials must be algebraically independent; secondly, they must satisfy the nullcone condition \cite{dk}. 

A set of polynomials $f_1, \dots, f_m \in K[x_1, \dots, x_k]$ is said to be {\em algebraically independent} if the only polynomial $h \in K[z_1,\dots, z_m]$ satisfying $h(f_1, \dots, f_m) = 0$ is the zero polynomial. Although trivial to define, the algebraic independence of polynomials is less trivial to check. One method proceeds via calculation of a Groebner basis, while another uses the Jacobi criterion. The former quickly becomes inefficient when used with many polynomials of high degree, but more importantly, it is difficult to apply in an abstract way. We therefore resort to using the Jacobi criterion\footnote{For proof of the Jacobi criterion, see for example \cite{Jacobi1} or \cite{Jacobi2}.} which states that a set of polynomials, $f_1, \dots, f_m \in K[x_1, \dots, x_k]$, is algebraically independent if and only if the wedge product of the exterior derivatives\footnote{The definition of a derivative requires some care for fields where limits are not defined \cite{Jacobi2}, but here we will only need to consider the case $K = \C$.} of the polynomials is non-zero, {\em i.e.}
\begin{align*}
df_1 \wedge \dots \wedge df_m \neq 0.
\end{align*}
As regards the nullcone condition, the {\em nullcone}, $N_V \subseteq V$, of an algebra $K[V]^G$ is defined to be the vanishing locus of all homogeneous invariant polynomials of strictly positive degree. That is,
\begin{align*}
N_V = \{ v \in V \ | \ f(v) = 0, \  \forall f \in K[V]_+^G \}.
\end{align*}
A set of polynomials, $\{f_1, \dots, f_m\}$, is said to satisfy the nullcone condition if the vanishing locus of all of its constituent polynomials coincides with $N_V$. We remark that, in the case of the Lorentz- and permutation-invariant algebra $\C[V]^{O(d) \times P}$, the nullcone is the set $\{ p_i \cdot p_j = 0, \forall i \leq j \}$. The fact that this does not depend on the choice of $P$ will prove to be important when we come to construct an HSOP for arbitrary $P$.

\subsection{A HSOP in $n=d+1$ with $P = 1$} \label{sec:HSOPSP=1}

Let us warm up by considering the case without permutations. With $n = d+1$, the SFT tells us that the relations between the dot products $p_i \cdot p_j$ are generated by the image of a single element under the Weyl map $W$, namely the determinant of the matrix whose $ij$-th entry is $y_{ij}$ for $i \leq j$ and $y_{ji}$ for $i > j$. Thus, $W(\mathrm{det}(y_{ij})) = \mathrm{det}(p_i \cdot p_j) = 0$ where $\mathrm{det}(p_i \cdot p_j) \in \C[V]^{O(d)}$. This will be important for proving that our proposed HSOP satisfies the nullcone condition. We now make the following

\begin{proposition} A HSOP for the algebra $\C[V]^{O(d)}$, with $n = d+1$, is given by the\\ $d(d+3)/2$ polynomials
\begin{equation} \label{eq:HSOPP=1}
\begin{aligned}
\theta_i \ &= p_1 \cdot p_1 + p_i \cdot p_i, \ \ \ 2 \leq i \leq d+1, \\
\alpha_{ij} &= p_i \cdot p_j, \hspace{1.5cm} 1 \leq i < j \leq d+1 .
\end{aligned}
\end{equation}
\end{proposition}

\textit{Proof:} We first check that these polynomials satisfy the nullcone condition. Evidently, if all dot products vanish, then both $\theta_i$ and $\alpha_{ij}$ vanish. Proceeding in the other direction, suppose that $\theta_i$ and $\alpha_{ij}$ vanish. The vanishing of $\alpha_{ij}$ implies not only the vanishing of the dot products with $i<j$, but also implies, together with the vanishing determinant relation, that $\prod_{i=1}^{d+1} (p_i \cdot p_i) = 0$. So either $(p_1 \cdot p_1) = 0$ or $(p_k \cdot p_k) = 0$ for some $2 \leq k \leq d+1$. If the former, then the fact that $\theta_i =0$ implies $p_i \cdot p_i=0$. If the latter, then $\theta_k = 0$ implies $p_1 \cdot p_1=0$, while the vanishing of the other $\theta_i$ implies the vanishing of all other $p_i \cdot p_i$ with $i \neq 1,k$. Either way, all dot products vanish and the nullcone condition is satisfied.

To prove algebraic independence, it is sufficient to show that the wedge product of the exterior derivatives of $\theta_i$ and $\alpha_{ij}$ is non-zero on at least a single point. We choose to evaluate the wedge product at the point
\begin{align*}
& p_1 = (0,0, 0, \dots, 0), \\ 
& p_2 = (1,0, 0, \dots, 0), \\ 
& p_3 = (0,1,0,\dots, 0), \\
& \vdots \\
& p_{d+1} = (0,0, \dots, 0, 1), \\ 
\end{align*}
where the unit entry moves progressively along, as indicated. We claim that the component of the wedge product proportional to
\begin{align*}
\omega = dp_1^1 \wedge \dots \wedge dp_1^d \wedge dp_2^1 \wedge \dots \wedge dp_2^d \wedge dp_3^2 \wedge \dots \wedge dp_3^d \wedge dp_4^3 \wedge \dots \wedge dp_4^d \wedge \dots \wedge dp_d^{d-1} \wedge dp_d^d \wedge dp_{d+1}^d,
\end{align*}
has coefficient at this point given by $2^{d} \neq 0$ (up to an irrelevant minus sign) and so the Jacobi criterion is satisfied. To establish the claim in detail, one starts by showing that the only non-zero contribution to the wedge product is
\begin{align*}
d(p_2 \cdot p_2) & \wedge \dots \wedge d(p_{d+1} \cdot p_{d+1})  \wedge d(p_1 \cdot p_2) \wedge \dots \wedge d(p_1 \cdot p_{d+1}) \\ 
& \wedge d(p_2 \cdot p_3) \wedge \dots \wedge d(p_2 \cdot p_{d+1}) \wedge d(p_3 \cdot p_4) \wedge \dots \wedge d(p_{d} \cdot p_{d+1}),
\end{align*}
as contributions with more than one $d(p_1 \cdot p_1)$ vanish trivially and contributions with a single $d(p_1 \cdot p_1)$ vanish on the specified point as $d(p_1 \cdot p_1) = 2 \sum_i p_1^i dp_1^i = 0$ there. Now, the coefficient of the component proportional to $\omega$ can be thought of as the determinant of an associated matrix.\footnote{Explicitly, it is the matrix with $ij$-th entry being $\frac{\partial f_i}{\partial x_j}$ where $f_i \in \{p_r \cdot p_s \ | \ r \leq s, \ s \neq 1\}$ and $x_j \in \{ p_1^1, \dots, p_1^d, p_2^1, \dots, p_2^d, p_3^2, \dots, p_3^d, p_4^3, \dots, p_4^d, \dots, p_d^{d-1}, p_d^d, p_{d+1}^d\}$.} In that form, after some row and column swaps (hence the irrelevant minus sign), one can show that the coefficient is the determinant of a diagonal matrix whose entries are all $1$'s except for $d$ instances of $2$'s which come from the $d(p_i \cdot p_i) = 2 \sum_j p_i^j dp_i^j = 2 dp_i^{i-1}$, for $2 \leq i \leq d+1$. Hence, the coefficient of $\omega$ is $2^d$ as claimed. 

Therefore, the proposed set of polynomials $\theta_i$ and $\alpha_{ij}$ satisfies the algebraic independence and nullcone conditions and so constitutes a valid HSOP. $\hfill \Box$

\subsection{A HSOP in $n=d+1$ with $P = S_n$} \label{sec:HSOPSP=Sn}

We now move on to the full-permutation case. Previously, in the $n = d+1$ case with no permutations, the SFT indicated that the relations between the dot products are generated by a single element, $\mathrm{det}(p_i \cdot p_j) \in \C[V]^{O(d)}$, where $\mathrm{det}(p_i \cdot p_j) = 0$. Here, we consider the full permutations case and work in the permutation-invariant subalgebra $\C[V]^{O(d) \times S_n}$. But, since the determinant relation, $\mathrm{det}(p_i \cdot p_j)$, is permutation-invariant, it is also an element of the permutation-invariant subalgebra, $\mathrm{det}(p_i \cdot p_j) \in \C[V]^{O(d) \times S_n}$, and therefore can be safely used in our proof of the validity of the HSOP for $P = S_n$. 

As to the HSOP itself, we take inspiration from Gauss who tells us that the $m$ symmetric polynomials in $m$ independent variables satisfy the necessary HSOP conditions. Therefore, the obvious candidates in our case are given by symmetric polynomials in the $d+1$ variables $p_i\cdot p_i$ and $d(d+1)/2$ variables $p_i\cdot p_j$ (with $i<j$), giving a total of $d(d+3)/2+1$. But, these cannot satisfy the algebraic independence condition since the dot products are not independent variables. It is therefore reasonable to suppose that in order to fix this, we need to judiciously discard one symmetric polynomial from this set to obtain a valid HSOP. As we will see, taking the power sum polynomials and discarding the highest degree polynomial in $p_i\cdot p_i$ does the job. In fact, taking any set of symmetric polynomials (elementary\footnote{The $k$-th elementary symmetric polynomial, $e_k$, on the the variables $x_1,\dots,x_n$ is defined as 
\begin{align*}
e_k(x_1,\dots,x_n)=\sum_{1\leq j_1 < j_2<\dots < j_k \leq n} x_{j_1} \dots x_{j_k}.
\end{align*}} or complete homogeneous) and discarding the highest degree polynomial in $p_i\cdot p_i$ also does the job. This can be seen by using Newton's identities for the elementary symmetric polynomials or the equivalent relations for the complete homogeneous symmetric polynomials.\footnote{Newton's identities relate the $m$-th power sum symmetric polynomial, $\mathrm{Pow}_m$, to the first $m$ elementary symmetric polynomials, $e_i$, via:
\begin{align*}
\mathrm{Pow}_m = \sum_{r_1 +2r_2 + \dots + m r_m = m, \atop r_1 \geq 0, \dots, r_m \geq 0} (-1)^m \frac{m (r_1 + \dots + r_m - 1)!}{r_1!r_2! \dots r_m!} \prod_{i=1}^m (-e_i)^{r_i}.
\end{align*}
Equivalent relations relating the power sum symmetric polynomials to the complete homogeneous symmetric polynomials also exist.
} Indeed, we  have the following

\begin{proposition} A HSOP for the algebra $\C[V]^{O(d) \times S_{n}}$, with $n = d+1$, is given by the $d(d+3)/2$ permutation-invariant polynomials
\begin{equation}\label{eq:hsopb}
\begin{aligned}
&\theta_k = \mathrm{Pow}_k(p_i \cdot p_i) := \sum_{i=1}^{d+1} (p_i \cdot p_i)^k, \ \ 1 \leq k \leq d, \\
&\alpha_k = \mathrm{Pow}_k(p_i \cdot p_j) :=  \sum_{i < j}^{d+1} (p_i \cdot p_j)^k, \ \ 1 \leq k \leq d(d+1)/2, 
\end{aligned}
\end{equation}
where $\mathrm{Pow}_k$ is the $k$-th power symmetric polynomial.
\end{proposition} 

\textit{Proof:} We first check that these polynomials satisfy the nullcone condition. Evidently, if all the dot products vanish, then both $\theta_k$ and $\alpha_k$ vanish. Proceeding in the other direction, suppose that $\theta_k$ and $\alpha_{k}$ vanish. Using Newton's identities, one can show that the vanishing of the first $r$ power symmetric polynomials implies the vanishing of the first $r$ elementary symmetric polynomials. Therefore, $\{ \alpha_{k} = 0, \  \forall k \}$ implies the vanishing of all the elementary symmetric polynomials on the $p_i \cdot p_j, i < j$. Now, the vanishing of the highest degree elementary symmetric polynomial, $\prod_{i<j}^n p_i \cdot p_j = 0$, implies the vanishing of at least one $p_i \cdot p_j,  \ i < j$. This then implies the vanishing of the $d(d+1)/2 - 1$ elementary symmetric polynomials on the remaining $d(d+1)/2 - 1$ dot products $p_i \cdot p_j, i < j$. Repeating this process recursively, one sees that the vanishing of the $\alpha_k$ implies the vanishing of all $p_i \cdot p_j, i < j$. This result, combined together with $\mathrm{det}(p_i \cdot p_j) = 0$, implies that $\prod_{i=1}^{d+1} (p_i \cdot p_i) = 0$. But $\prod_{i=1}^{d+1} (p_i \cdot p_i)$ is the elementary symmetric polynomial of highest degree, so $\prod_{i=1}^{d+1} (p_i \cdot p_i) = 0$, together with the vanishing of the $ \theta_k $, implies the vanishing of all $d+1$ elementary symmetric polynomials in $p_i \cdot p_i$. From here, one can again recursively show that all $p_i \cdot p_i$ must vanish, so the nullcone condition is satisfied.

To prove algebraic independence, we evaluate (a component of) the wedge product of the exterior derivatives of $\theta_k$ and $\alpha_k$ at the point
\begin{align*}
& p_1 = (2,0, \dots, 0), \\ 
& p_2 = (3,0, \dots, 0), \\ 
& p_3 = (l_{m},1, 0, \dots, 0), \\ 
& p_4 = (l_{m+1},0, 1, 0, \dots, 0), \\ 
& \vdots \\ 
& p_i = (l_{m+i-3},0, \dots,0, 1, 0, \dots, 0), \\ 
& \vdots \\ 
& p_{d+1} = (l_{m+d-2},0, \dots, 0, 1), \\ 
\end{align*}
where $l_{i}$ denotes the $i$-th prime number (with $l_1 = 2$) and $m \geq 3$ and where the unit entry moves progressively along, as indicated. The prime numbers will prove useful soon when we require that the dot products $p_i \cdot p_j$ are all distinct. We claim that the component of the wedge product proportional to
\begin{align*}
\omega = dp_1^1 \wedge \dots \wedge dp_1^d \wedge dp_2^1 \wedge \dots \wedge dp_2^d \wedge dp_3^2 \wedge \dots \wedge dp_3^d \wedge dp_4^3 \wedge \dots \wedge dp_4^d \wedge \dots \wedge dp_d^{d-1} \wedge dp_d^d \wedge dp_{d+1}^d,
\end{align*}
has a non-zero coefficient. To establish this claim in detail, we first note that the wedge product can be re-expressed as 
\begin{align*}
d! \left( \frac{d(d+1)}{2} \right)! \mathrm{det}(M) \sum_{k=1}^{d+1} \mathrm{det}(L_k) \Omega_k,
\end{align*}
where $M$ is the Vandermonde matrix\footnote{The Vandermonde matrix $V$ on a set of variables $x_i$, $i \in \{1,\dots,n\}$, is the $n \times n$ matrix with entries $V_{ij} = x_i^{j-1}$. The determinant of this matrix can be nicely expressed as $\mathrm{det}(V) = \prod_{1\leq i < j \leq n} (x_i - x_j)$ and is non-zero only if all the $x_i$'s are distinct.} on the $p_i \cdot p_j, i<j$, $L_k$ is the Vandermonde matrix on the $p_i \cdot p_i, i \neq k$, and 
\begin{align*}
\Omega_k = d(p_1 \cdot p_1) \wedge \dots \wedge \widehat{d(p_k \cdot p_k)} \wedge \dots \wedge  d(p_{d+1} \cdot p_{d+1})  \wedge d(p_1 \cdot p_2) \wedge d(p_1 \cdot p_3) \wedge \dots \wedge d(p_{d} \cdot p_{d+1}),
\end{align*} 
where $ \ \widehat{} \ $ over a term indicates that that term should be omitted. By considering the coefficient of the component proportional to $\omega$ of $\Omega_k$ as the determinant of an associated matrix\footnote{Explicitly, it is the matrix with $ij$-th entry being $\frac{\partial f_i}{\partial x_j}$ where $f_i \in \{p_r \cdot p_s \ | \ r \leq s, \ s \neq k\}$ and $x_j \in \{ p_1^1, \dots, p_1^d, p_2^1, \dots, p_2^d, p_3^2, \dots, p_3^d, p_4^3, \dots, p_4^d, \dots, p_d^{d-1}, p_d^d, p_{d+1}^d\}$.}, one can show that the only contributions to the sum on the specified point come from the instances with $k = 1, 2$. Therefore, the coefficient of the component proportional to $\omega$ of the wedge product is
\begin{align*}
2^{d} d! \left( \frac{d(d+1)}{2} \right)! \mathrm{det}(M) \left( 9 \mathrm{det}(L_1)  - 4 \mathrm{det}(L_2) \right),
\end{align*}
up to an irrelevant overall minus sign (from row and column swaps). The $\mathrm{det}(M)$ term is non-zero as every dot product $p_i \cdot p_j$, $i<j$, is distinct (our use of prime numbers guarantees that $l_{i} l_{j} = l_{m}l_{n}$ if and only if either $l_i = l_n$ and $l_j = l_m$ or $l_i = l_m$ and $l_j = l_n$). To show the last term is non zero, we expand it as
\begin{align*}
&\left( 9 \mathrm{det}(L_1)  - 4 \mathrm{det}(L_2) \right)  = 9 \prod_{i<j \neq 1}^{d+1} (p_i \cdot p_i - p_j \cdot p_j) - 4 \prod_{i<j \neq 2}^{d+1} (p_i \cdot p_i - p_j \cdot p_j) \\
& = \prod_{i < j \neq 1,2}^{d+1} (p_i \cdot p_i - p_j \cdot p_j) \left( 9 \prod_{i = 3}^{d+1} (p_2 \cdot p_2 - p_i \cdot p_i) - 4 \prod_{i =3}^{d+1} (p_1 \cdot p_1 - p_i \cdot p_i) \right).
\end{align*}
Since we have the freedom to choose $m$ to be as large as we want (there are infinitely many primes), we can see that this term is non-zero as follows: for large $m$, where $p_i \cdot p_i \gg p_1 \cdot p_1$, $p_2 \cdot p_2$, it tends to $\sim 5 \prod_{i = 3}^{d + 1} (p_i \cdot p_i) \prod_{i < j \neq 1,2}^{d+1} (p_i \cdot p_i - p_j \cdot p_j)$ which is non-zero as the $p_i \cdot p_i$ are non-zero and distinct (again by our use of the prime numbers). Therefore, the Jacobi criterion is satisfied.

Hence, these polynomials are algebraically independent and satisfy the nullcone condition and so constitute a valid HSOP. $\hfill \Box$

As we have already remarked, the nullcone of $\C[V]^{O(d) \times P}$, being given by the vanishing locus of the dot products $p_i\cdot p_j$, is independent of the choice of $P$. Moreover, an algebraically independent set of $O(d,\C) \times S_{n}$-invariant polynomials is also an algebraically independent set of $O(d,\C) \times P$-invariant polynomials, for any $P \subset S_n$. We thus have the important
\begin{corollary} A HSOP for the algebra $\C[V]^{O(d) \times P}$, with $n = d+1$, is given by the permutation-invariant polynomials in \ref{eq:hsopb}, for any $P \subset S_n$.
\end{corollary}

As we shall see, this gives us a starting point for finding a Hironaka decomposition for any $P$ in the case $n=d+1$.

\subsection{A remark on HSOPs for $n \geq d + 2$}
It would obviously be desirable to generalise our methods to cases with $n \geq d + 2$.  The first obstacle in doing so is that the relations between the dot products $p_i \cdot p_j$ given by the higher minors of the matrix whose entries are $p_i \cdot p_j$, are not $S_n$-invariant. Thus, they do not belong to 
$\C[V]^{O(d) \times S_n}$ and cannot be used directly in the proofs.  To overcome this, one presumably needs to first find a set of invariant polynomials which generate the relations and then work with these. But it is not clear to us what form a HSOP might take.

\section{Secondaries} \label{sec:Secs} 
Now that we can write down HSOPs of our invariant algebras at will in cases with $n \leq d+1$, the corresponding secondaries may be computed via an algorithm (which can be found in \cite{dk}). In this Section, we illustrate the algorithm by applying it to a simple example, namely $(n,d) = (3,2)$ with no permutation symmetry, {\em i.e.} the algebra $\C[V]^{O(2)}$, with $V \cong \C^6$.

The algorithm is based on the following two observations. Firstly, the number of secondaries required can be read off (along with their degrees) from the Hilbert series, which itself can be computed using standard methods from invariant theory (as we review in the Appendix).  Indeed, given a Hironaka decomposition of an invariant algebra $K[V]^G = \bigoplus_i \eta_i K[\theta_j]$, its Hilbert series $H(K[V]^G,t)$, takes the form $\frac{1 + \sum_{k=1} S_k t^{k}}{\prod_{l=1} (1-t^l)^{P_l}}$ where $S_k$ is the number of secondary invariants $\eta_i$ of degree $k$ and $P_l$ is the number of primary invariants (HSOP) $\theta_j$ of degree $l$. Therefore, given a HSOP, which fixes the $P_l$, one can read off the number and degrees of the secondaries from the numerator of the Hilbert series. Secondly, given a set of polynomial invariants $\{ \eta_1, \dots, \eta_m\}$ of the right cardinality, the set forms the secondaries of the invariant algebra if and only if its constituent polynomials are linearly independent modulo the ideal $I := \langle \theta_1,\dots, \theta_r \rangle \in K[V]$ generated by the HSOP $\{ \theta_i \}$. To show linear independence of a set of polynomials modulo an ideal, one can compute the remainders of the polynomials upon division by a Groebner basis of that ideal and check that the remainders are themselves linearly independent \cite{dk}.

Turning to our example, the methods described in the Appendix show that the Hilbert series is given by
\begin{align*}
H(\C[V]^{O(2)}, t) = \frac{1 + t^2 + t^4}{(1-t^2)^5}.
\end{align*}
Here we have written the series in a form such that the denominator reproduces the five primaries of degree 2 corresponding to the HSOP given in Equation \ref{eq:HSOPP=1}, namely
\begin{align*}
\{ (p \cdot p) + (q \cdot q),  (p \cdot p) + (r \cdot r) ,  (p \cdot q),  ( p \cdot r),  (q \cdot r) \},
\end{align*}
where we have labelled the momenta by $p,q,$ and $r$ (we denote the corresponding components of $V$ by $\{p_1, p_2, q_1, q_2, r_1, r_2 \}$). We thus read off from the numerator that there is $1$ secondary of degree $2$ and $1$ secondary of degree $4$ (and of course the trivial secondary, $1$, of degree $0$).

The next step in the algorithm is to compute a Groebner basis of the ideal generated by the HSOP, which will later be used to verify the linear independence of the secondaries. To do so, one must first choose a monomial ordering.\footnote{Readers seeking a gentle introduction to Groebner basis methods may wish to consult \cite{CoxLittleOShea}.} A common (and often very efficient) choice is graded reverse lexicographic order.\footnote{Graded reverse lexicographic order, or grevlex for short, is a monomial ordering on some variables $x_1,\dots,x_n$ where for any two monomials $t = x_1^{a_1} \dots x_n^{a_n}$ and $t^\prime = x_1^{a^\prime_1} \dots x_n^{a^\prime_n}$, $t >_{\mathrm{grevlex}} t^\prime$ if $\mathrm{deg}(t) > \mathrm{deg}(t^\prime)$ or if $\mathrm{deg}(t) = \mathrm{deg}(t^\prime)$ and $a_i < a^\prime_i$ for the largest $i$ with $a_i \neq a^\prime_i$.} In this ordering, a Groebner basis of the ideal generated by our HSOP is given by the set of $20$ polynomials
\begin{align*}
\mybox{ \{q_1 r_1+q_2 r_2,p_1 r_1+p_2 r_2,q_1^2+q_2^2-r_1^2-r_2^2,p_1 q_1+p_2 q_2,p_1^2+p_2^2+r_1^2+r_2^2,p_2 q_1 r_2-p_1 q_2 r_2,q_2^2 r_1-q_1 q_2
   r_2-r_1^3-r_2^2 r_1,p_2 q_2 r_1-p_1 q_2 r_2,p_2^2 r_1-p_1 p_2 r_2+r_1^3+r_2^2 r_1,-p_1 q_2^2+p_2 q_1 q_2+p_1 r_2^2-p_2 r_1 r_2,p_2^2 q_1-p_1 p_2
   q_2+q_1 r_2^2-q_2 r_1 r_2,q_2 r_1 r_2^2-q_1 r_2^3,p_2 r_1 r_2^2-p_1 r_2^3,r_2 r_1^3+r_2^3 r_1,q_2 r_2^3+q_2 r_1^2 r_2,p_2 r_2^3+p_2 r_1^2
   r_2,r_1^4-r_2^4,q_2 r_1^3+q_1 r_2^3,p_2 r_1^3+p_1 r_2^3,r_2^5+r_1^2 r_2^3 \}.}
\end{align*}
We then proceed to generate a basis of homogeneous invariant polynomials in the algebra of degree $d_i$, corresponding to the degrees of secondaries read off of the Hilbert series, using linear algebra methods. If $G$ were a finite group, this would be a simple matter of averaging all possible monomials of degree $d_i$ over $G$ to obtain a basis of invariant polynomials at that degree.\footnote{The average of a polynomial $f$ over a finite group $G$ is $\frac{1}{|G|} \sum_{g \in G} g \circ f$.} But for us $G$ is infinite, so things are not so straightforward. We use the additional information that $\C[V]^{O(d) \times P} \subset \C[V]^{O(d)}$ and that by the FFT, $\C[V]^{O(d)}$ is generated by the set of dot products in the momenta. This allows one to obtain a basis of homogeneous polynomials in $\C[V]^{O(d) \times P}$ of degree $d_i$ by averaging all possible products of $d_i / 2$ dot products over the (finite) permutation group $P$.

From this basis, we consecutively choose elements and compute their remainders upon division by the Groebner basis (also called the normal forms) and keep them only if their remainders are non-zero and lie outside the $\C$-vector space generated by the remainders of previously found secondaries ({\em i.e.} the remainders are linearly independent). Once the required number of secondaries is obtained, one proceeds to the next degree and so on until all the secondaries have been found. 

In our case (skipping over the trivial case of the secondary $1$), we start at degree $2$. Here, the basis of polynomials is just the set of dot products. Choosing $p \cdot p$, we compute the remainder upon division to be $-(r_1^2+r_2^2)$, which is non-zero and so we have the required secondary of degree $2$. We then move on to degree $4$. Here, the basis of polynomials is all possible products of two dot products. We choose $(p \cdot p)^2$ and compute the remainder upon division to be $2 r_2^2(r_1^2 +r_2^2)$, which is non-zero and is obviously linearly independent from the remainder of the previous secondary, since it does not have the same degree. We therefore have our required secondary of degree $4$.\footnote{We could have equally chosen either $(q \cdot q)$ or $(r \cdot r)$ for the degree $2$ secondary and either $(q \cdot q)^2$ or $(r \cdot r)^2$ for the degree $4$ one. It is also interesting to note that the remainders upon division by the Groebner basis of $(p \cdot p)^3,(q \cdot q)^3,$ and $(r \cdot r)^3$ are zero and so they lie {\em in} the ideal generated by the HSOP.} Finally, we obtain the Hironaka decomposition of the algebra as follows 
\begin{align} \label{eq: HK (3,2)}
\C[V]^{O(2)} = \left( 1 \oplus (p\cdot p)  \oplus (p \cdot p)^2 \right) \cdot \C[p \cdot p + q \cdot q,  p \cdot p + r \cdot r, p \cdot q,   p \cdot r,  q \cdot r].
\end{align}

Simple though it is, our example already hints at the two bottlenecks that arise when computing the secondaries of $\C[V]^{O(d) \times P}$ in high dimensions with large permutation symmetry. One is the computation of the Groebner basis of the ideal and the other is the computation of a basis of invariant polynomials of a certain degree, which becomes progressively more costly at higher degrees. There are multiple tricks which can be used to mitigate the latter bottleneck \cite{dk} ({\em e.g.}, using products of lower degree secondaries as candidates), but there is still no really effective way of tackling the inefficiency of the Groebner basis computations. 

In the next Section, we employ the algorithm to provide Hironaka decompositions for computationally tractable cases. A version of this algorithm is implemented in \texttt{Macaulay2} \cite{M2}, amongst others. 

\section{Examples} \label{sec:Examples}

We now present two examples with explicit Hironaka decompositions using the above prescriptions.

\subsection{The case of $(n,d) = (5,4)$ with $P = 1$}

For the no permutation case with $(n,d) = (5,4)$, we start by finding the HSOP for the algebra $\C[V]^{O(4)}$ in the way described in Section \ref{sec:HSOPSP=1}. This results in the following set of polynomials
\begin{align*}
&\theta_i = p_1 \cdot p_1 + p_i \cdot p_i, \ \ 2 \leq i \leq 5, \\
&\alpha_{ij} = p_i \cdot p_j, \ \ 1 \leq i < j \leq 5.
\end{align*}
Using the algorithm described in Section \ref{sec:Secs}, we proceed to find the secondaries in a similar manner. The Hilbert series of the algebra, computed using methods described in the Appendix, is 
\begin{align*}
H(\C[V]^{O(4)}, t) = \frac{1 + t^2 + t^4 + t^6 + t^8}{(1-t^2)^{14}}.
\end{align*} 
We are therefore looking for $1$ secondary at each of the degrees $0,2,4,6,$ and $8$. We find that the following set of polynomials 
\begin{align*}
1, (p_1 \cdot p_1), (p_1 \cdot p_1)^2, (p_1 \cdot p_1)^3, (p_1 \cdot p_1)^4,
\end{align*}
have remainders upon division by the Groebner basis of the ideal generated by the HSOP which are non-zero and linearly independent. Therefore, we obtain a Hironaka decomposition of the algebra as follows 
\begin{align*}
\C[V]^{O(4)} = \left( 1 \oplus (p_1 \cdot p_1)  \oplus (p_1 \cdot p_1)^2  \oplus (p_1 \cdot p_1)^3\oplus (p_1 \cdot p_1)^4 \right) \cdot \C[\{\theta_i, \alpha_{ij}\}].
\end{align*}

The forms of the Hironaka decompositions for $d=2$ and $4$ given here and in \ref{eq: HK (3,2)} invite an obvious conjecture for their form in arbitrary dimension $d$. Namely, the secondaries are given by the dot product of any one momenta with itself, raised to the $0$-th all the way to the $d$-th powers. An explicit computation shows this to be the case also in $d=1$ and 3. Therefore, we are led to the following
\begin{conjecture} 
The Hironaka decomposition of Lorentz-invariant algebras, $\C[V]^{O(d)}$, in the case of $n = d + 1$, is given by
\begin{align*}
\C[V]^{O(d)} = \bigoplus_{m = 0}^{d} (p_1 \cdot p_1)^m \ \C[\{\theta_i, \alpha_{ij}\}],
\end{align*}
where the HSOP $\{\theta_i, \alpha_{ij}\}$ are as given by Equation \ref{eq:HSOPP=1}.
\end{conjecture}

\subsection{The case of $(n,d) = (3,2)$ with $P = S_3$}
For the full permutation case with $(n,d) = (3,2)$, we find the HSOP for the algebra $\C[V]^{O(2) \times S_3}$ in the way described in Section \ref{sec:HSOPSP=Sn}. This results in the following set
\begin{align*}
&\theta_k = \mathrm{Pow}_k(p_i \cdot p_i) = \sum_{i=1}^{3} (p_i \cdot p_i)^k, \ \ 1 \leq k \leq 2,  \\
&\alpha_k = \mathrm{Pow}_k(p_i \cdot p_j) = \sum_{i < j}^{3} (p_i \cdot p_j)^k, \ \ 1 \leq k \leq 3.
\end{align*}
Using the algorithm described in Section \ref{sec:Secs}, we proceed to find the secondaries in a similar manner. The Hilbert series of the algebra, computed using methods described in the Appendix, is 
\begin{align*}
H(\C[V]^{O(2) \times S_3}, t) =  \frac{1 + t^4 + 2t^6 + t^8 + t^{12}}{\left(1-t^2\right)^2 \left(1-t^4\right)^2 \left(1-t^6\right)}.
\end{align*} 
We are therefore looking for $1$ secondary at degree $0$, $1$ at degree $4$, $2$ at degree $6$, 1 at degree $8$, and $1$ at degree $12$. We find that the following set of polynomials 
\begin{align*}
& \eta_1 = 1, \\
& \eta_2 = (p_1 \cdot p_1)(p_2 \cdot p_3) + (p_2 \cdot p_2)(p_1 \cdot p_3) + (p_3 \cdot p_3)(p_1 \cdot p_2), \\
& \eta_3 = (p_1 \cdot p_1)^2(p_2 \cdot p_3) + (p_2 \cdot p_2)^2(p_1 \cdot p_3) + (p_3 \cdot p_3)^2(p_1 \cdot p_2), \\
& \eta_4 = (p_1 \cdot p_1)(p_2 \cdot p_3)^2 + (p_2 \cdot p_2)(p_1 \cdot p_3)^2 + (p_3 \cdot p_3)(p_1 \cdot p_2)^2, \\
& \eta_5 = (p_1 \cdot p_1)^2(p_2 \cdot p_3)^2 + (p_2 \cdot p_2)^2(p_1 \cdot p_3)^2 + (p_3 \cdot p_3)^2(p_1 \cdot p_2)^2, \\
& \eta_6 = (p_1 \cdot p_1)^5(p_2 \cdot p_3) + (p_2 \cdot p_2)^5(p_1 \cdot p_3) + (p_3 \cdot p_3)^5(p_1 \cdot p_2), 
\end{align*}
have remainders upon division by the Groebner basis of the ideal generated by the HSOP which are non-zero and linearly independent. Therefore, we obtain a Hironaka decomposition of the algebra as follows 
\begin{align*}
\C[V]^{O(2) \times S_3} = \bigoplus_{i=1}^6 \eta_i \C[\{\theta_k, \alpha_k\}].
\end{align*}

\section{Discussion}

In this work, we have addressed the problem of redundancies in the description of the Lorentz- and permutation-invariant algebras via generating sets. Instead of providing a set of generators (FFT) and the relations between them (SFT), we observed that one may provide (via the theorem of Hochster and Roberts) a more direct characterization in terms of a Hironaka decomposition, that is as a free, finitely-generated module over a polynomial subalgebra. In cases where $n \leq d+1$, we gave an explicit solution (for an arbitrary permutation symmetry) to the `hard' part of finding such a decomposition, namely the identification of a homogeneous system of parameters (HSOP). The `easy' part of finding a decomposition, namely the identification of suitable secondary generators, reduces to a linear algebra algorithm, but is nonetheless inefficient. 
We provided Hironaka decompositions in the examples of $(n,d) = (5,4)$ with $P = 1$ and $(n,d) = (3,2)$ with $P = S_3$ and a conjecture in the general case of $n = d + 1$ with no permutations.

\appendix
\addcontentsline{toc}{section}{\hspace{-1ex}Appendix: Hilbert Series of $\C[V]^{O(d) \times P}$}
\addtocontents{toc}{\protect\setcounter{tocdepth}{-1}}
\section*{Appendix: Hilbert Series of $\C[V]^{O(d) \times P}$} \label{app:HS}

In this Appendix, we describe how to compute Hilbert series of invariant algebras under the combined (complexified) Lorentz and permutation groups in dimension $2 \leq d \leq 4$.

To do so, we use a generalisation of Molien's formula valid for a reductive group $G$, whereby the Hilbert series of an invariant algebra $\C[V]^G$ is given by \cite{dk}
\begin{align*}
H(\C[V]^G,t) = \int_C \frac{d\mu}{\mathrm{det}_V (1 - t\cdot \rho_V)},
\end{align*}
where $C$ is a maximal compact subgroup of $G$, $d\mu$ is a Haar measure on $C$ normalised such that $ \int_C d\mu = 1$, and $\rho_V : C \rightarrow GL(V)$ denotes the representation of $C$ carried by $V$. 
For what follows, it is useful to note that the integrand is constant within a conjugacy class of $G$.

We now consider in turn the cases of $d = 2,3,$ and $4$, with an arbitrary number of momenta $n$ and an arbitrary permutation group, $P \subset S_n$ acting on those momenta. The complexification of the Lorentz group when parity is a symmetry means that the groups we consider are of the form $G = O(d,\C) \times P$. For completeness, we also discuss the case where parity is not a symmetry, {\em i.e.} when $G = SO(d,\C) \times P$.

\subsection*{The case of $O(2,\C) \times P$}

We start by considering the invariant algebra $\C[V]^G$ in the case of $n$ momenta in $2$ dimensions with no permutation symmetry which corresponds to $G = O(2,\C)$ and $V \cong \C^{2n}$. The group $O(2,\C)$ has maximal compact subgroup $O(2,\R) \cong U(1) \rtimes \Z_2$ and its action on $\C^2$ may be written as\footnote{If we consider $O(2,\R) \subset O(2,\C)$ as acting on the real components of the momenta, then the isomorphism $O(2,\R) \cong U(1) \rtimes \Z_2$  corresponds to the linear map $(p_0,p_1) \in \C^2 \mapsto (p_0 + i p_1, p_0 - i p_1)$.}
\begin{align*}
 M^+(z) = \begin{pmatrix} z & 0 \\ 0 & z^{-1} \end{pmatrix},
M^-(z) = \begin{pmatrix} 0 & z^{-1} \\ z & 0 \end{pmatrix},  
\end{align*}
where $z \in \C$ such that $|z| = 1$ and where $M^+$ corresponds to the component connected to the identity and $M^-$ corresponds to the other connected component. When acting on $n$ copies of $\C^2$ (corresponding to $n$ particles), we have
\begin{align*}
M^\pm_V = 
\begin{pmatrix} 
M^\pm  &  &  &  \\ 
  & M^\pm  &  &  \\ 
 &  &  \ddots &  \\ 
&  &  & M^\pm
\end{pmatrix}.
\end{align*}
The normalised Haar measure is given by $\frac{1}{2} \frac{1}{2\pi i}\frac{dz}{z}$ on each component (which is half the Haar measure for the group $U(1)$ and so takes into account the 2 disconnected components). The Hilbert series is thus given by
\begin{align*}
H(\C[V]^{O(2)}, t) = \frac{1}{2} \frac{1}{2\pi i} \oint_{|z| = 1} \frac{dz}{z} \left( \frac{1}{\mathrm{det}_V (1 - t\cdot M^+_V)} + \frac{1}{\mathrm{det}_V (1 - t\cdot M^-_V)} \right).
\end{align*}  
For our example of $n = 3$ with $P = 1$, the integral becomes
\begin{align*}
H(\C[V]^{O(2)}, t) & = \frac{1}{2} \frac{1}{2\pi i} \oint_{|z| = 1} \frac{dz}{z} \left( \frac{1}{(1-t z)^3 (1-t/z)^3} + \frac{1}{(1-t^2)^3} \right) \\
& = \frac{1}{2} \left( \frac{1 + 4 t^2 + t^4}{(1 - t^2)^5} + \frac{1}{(1 - t^2)^3} \right) = \frac{1 + t^2 + t^4}{(1 - t^2)^5}.
\end{align*}
where the integrals have been carried out using the residue theorem of contour integration. 

We now include some permutation group $P \subseteq S_n$ acting on the $n$ momenta so that the combined group becomes $G = O(2,\C) \times P$ and its maximal compact subgroup is just $O(2,\R) \times P$. Here, one must additionally average over the permutation group $P$, where the action of $P$ simply permutes the $n$ particles, ergo the $n$ copies of $\C^2$. Since the integrand is constant within conjugacy classes, it suffices to pick one representative element from each class, and weight accordingly. The Haar measure is rescaled by $1/|P|$ so that it is still properly normalised.

For our example of $n=3$ with $P = S_3$, we have $3$ conjugacy classes: the identity with multiplicity 1, $(\cdot \cdot)$ with multiplicity 3, and $(\cdot \cdot \cdot)$ with multiplicity 2. We use the following representative elements from each permutation conjugacy class
\begin{align*}
\begin{pmatrix} 
M^\pm & 0  & 0 \\ 
0 & M^\pm & 0 \\ 
0 & 0 & M^\pm
\end{pmatrix},
\begin{pmatrix} 
0 & M^\pm & 0 \\ 
M^\pm & 0  & 0 \\ 
0 & 0 & M^\pm
\end{pmatrix},
\begin{pmatrix} 
0 & M^\pm & 0 \\ 
0 & 0 & M^\pm \\
M^\pm & 0  & 0
\end{pmatrix}.
\end{align*}
The contribution of the component connected to the identity then becomes
\begin{align*}
& H^+(\C[V]^{O(2) \times S_3}, t) \\
& = \frac{1}{6} \frac{1}{2} \frac{1}{2\pi i} \oint_{|z| = 1} \frac{dz}{z} \left( \frac{1}{(1-t z)^3 (1-t/z)^3} + \frac{3}{(1-t z) (1-t/z) (1-(t z)^2)(1-(t/z)^2)} + \frac{2}{(1-(t z)^3)(1-(t/z)^3)} \right)  \\
& = \frac{1}{2} \frac{1 + 3 t^4 + 4 t^6 + 3 t^8 + t^{12}}{(1-t^2)^2 (1-t^4)^2 (1-t^6)}.
\end{align*}
Similarly, the contribution of the other connected component is
\begin{align*}
H^-(\C[V]^{O(2) \times S_3}, t) =\frac{1}{2} \frac{1 + t^4}{(1 - t^2)^2 (1 - t^6)},
\end{align*}
and so finally we obtain 
\begin{align*}
H(\C[V]^{O(2) \times S_3}, t) &=  H^+(\C[V]^{O(2) \times S_3}, t) + H^-(\C[V]^{O(2) \times S_3}, t)  \\
& =  \frac{1 + t^4 + 2 t^6 + t^8 + t^{12}}{(1-t^2)^2 (1-t^4)^2 (1-t^6)}.
\end{align*}
Notice that we also get the Hilbert series for the case $G = SO(2, \C) \times P$, corresponding to when parity is not a symmetry, for free, by just considering the component connected to the identity
\begin{align*}
H(\C[V]^{SO(2)}, t) & = \frac{1 + 4 t^2 + t^4}{(1 - t^2)^5}, \\ 
H(\C[V]^{SO(2) \times S_3}, t) &= \frac{1 + 3 t^4 + 4 t^6 + 3 t^8 + t^{12}}{(1-t^2)^2 (1-t^4)^2 (1-t^6)}.
\end{align*}

\subsection*{The case of $O(3,\C) \times P$}
In $d=3$, the group $O(3,\C)$ has maximal compact subgroup $O(3, \R) \cong (SU(2) / \Z_2) \times \Z_2$. Since the integrand is constant on the conjugacy classes, we need consider only the maximal torus of $SU(2)$ with elements
\begin{align*}
\begin{pmatrix}
z & 0 \\
0 & z^{-1}
\end{pmatrix},
\end{align*}
where $|z| = 1$\footnote{Strictly speaking, one should consider only half of the unit circle, since $z$ and $-z$ yield the same element in $SU(2)/\Z_2$. But since the integral will turn out to be symmetric under $z \to -z$, we can get away with integrating over the whole circle.} acting on $\C^3$ as\footnote{Here, the isomorphism $O(3, \R) \cong (SU(2) / \Z_2) \times \Z_2$ corresponds to the linear map  $(p_0, p_1, p_2) \in \C^3 \mapsto (p_0 - i p_1, p_2, p_0 + i p_1)$.}
\begin{align*}
M^+(z) = 
\begin{pmatrix} 
z^2 & 0 & 0 \\ 0 & 1 & 0 \\ 0 & 0 & z^{-2}
\end{pmatrix},
M^-(z) = 
\begin{pmatrix} 
-z^2 & 0 & 0 \\ 0 & -1 & 0 \\ 0 & 0 & -z^{-2}
\end{pmatrix},
\end{align*}
where $\pm$ again distinguishes the 2 connected components. The normalised Haar measure on each component is $\frac{1}{2}\frac{1}{2 \pi i} \frac{(1-z^2)dz}{z}$ (which is just half of the usual normalised Haar measure for $SU(2)$). The Hilbert series with $n$ particles is then given by
\begin{align*}
H(\C[V]^{O(3)}, t) = & \frac{1}{2}\frac{1}{2\pi i} \oint_{|z| = 1} \frac{(1-z^2)dz}{z} \left( \frac{1}{\mathrm{det}_V (1 - t\cdot M^+_V)} + \frac{1}{\mathrm{det}_V (1 - t\cdot M^-_V)} \right).
\end{align*} 
For example, with $n = 4$ the integral becomes
\begin{align*}
H(\C[V]^{O(3)}, t) = & \frac{1}{2}\frac{1}{2\pi i} \oint_{|z| = 1}  \frac{(1-z^2)dz}{z} \left( \frac{1}{(1-t)^4 \left(1-\frac{t}{z^2}\right)^4 \left(1-t z^2\right)^4} + \frac{1}{(1+t)^4 \left(1+\frac{t}{z^2}\right)^4 \left(1+t z^2\right)^4} \right),
\end{align*} 
which we evaluate using the residue theorem, obtaining
\begin{align*}
H(\C[V]^{O(3)}, t) = \frac{1}{2} \left( \frac{1+t^2+4 t^3+t^4+t^6}{\left(1-t^2\right)^9} +\frac{1+t^2-4 t^3+t^4+t^6}{\left(1-t^2\right)^9} \right) = \frac{1+t^2+t^4+t^6}{\left(1-t^2\right)^9}.
\end{align*}
We also obtain the Hilbert series for when $G = SO(3,\C)$ for free by only considering the component connected to the identity
\begin{align*}
H(\C[V]^{SO(3)}, t) =  \frac{1+t^2+4 t^3+t^4+t^6}{\left(1-t^2\right)^9}.
\end{align*}
To include an arbitrary permutation group $P \subseteq S_n$ acting on the $n$ momenta, one needs to average over the conjugacy classes of $P$ as discussed previously.

\subsection*{The case of $O(4,\C) \times P$}
In $d = 4$, $O(4,\C)$ has maximal compact subgroup $O(4,\R) \cong \left((SU(2) \times SU(2))/ \Z_2 \right) \rtimes \Z_2$, where the automorphism in the semi-direct product corresponds to interchanging the 2 $SU(2)$ factors. Since the integrand is constant on the conjugacy classes, we need consider only the maximal torus with elements
\begin{align*}
\left(
\begin{pmatrix}
z & 0 \\
0 & z^{-1}
\end{pmatrix},
\begin{pmatrix}
w & 0 \\
0 & w^{-1}
\end{pmatrix}\right),
\end{align*}
where $|z| = |w| = 1$.\footnote{As in $d=3$, there is no need to take care in projecting to $(SU(2) \times SU(2))/ \Z_2$.} The action on $\C^4$ is given by\footnote{The asymmetry in the formul\ae\ arises from the fact that the conjugacy classes  in the disconnected component can be parameterized by a single $U(1)$; for details see \cite{TopPartners}.}\footnote{Here, the isomorphism $O(4, \R) \cong \left( (SU(2) \times SU(2))/ \Z_2 \right) \rtimes \Z_2$ corresponds to the linear map  $(p_0, p_1, p_2,p_4) \in \C^4 \mapsto (p_0 + i p_3, p_1 + i p_2, p_1 -i p_2, p_0 - i p_3)$.}
\begin{align*}
M^+(z,w) = 
\begin{pmatrix} 
z w & 0 & 0 & 0 \\ 0 & zw^{-1} & 0 & 0 \\ 0 & 0 & wz^{-1} & 0 \\ 0 & 0 & 0 & (z w)^{-1}
\end{pmatrix},
M^-(z) = 
\begin{pmatrix} 
z & 0 & 0 & 0 \\ 0 & 0 & z & 0 \\ 0 & z^{-1} & 0 & 0 \\ 0 & 0 & 0 & z^{-1}
\end{pmatrix},
\end{align*}
The normalised Haar measure on the component connected to the identity is $\frac{1}{2}\frac{1}{(2 \pi i)^2} \frac{(1-z^2)dz}{z}  \frac{(1-w^2)dw}{w}$ and the Haar measure on the disconnected component is $\frac{1}{2}\frac{1}{2 \pi i} \frac{(1-z^2)dz}{z}$. The Hilbert series with $n$ particles is then given by
\begin{align*}
H(\C[V]^{O(4)}, t) = & \frac{1}{2}\frac{1}{(2\pi i)^2} \oint_{|z| = |w| = 1} \frac{(1-z^2)(1-w^2)dzdw}{z w} \frac{1}{\mathrm{det}_V (1 - t\cdot M^+_V)} \\ 
& +  \frac{1}{2}\frac{1}{2\pi i} \oint_{|z| = 1} \frac{(1-z^2)dz}{z} \frac{1}{\mathrm{det}_V (1 - t\cdot M^-_V)} .
\end{align*} 
In our example of $n = 5$ with $P=1$, the integral becomes
\begin{align*}
H(\C[V]^{O(4)}, t) = & \frac{1}{2}\frac{1}{(2\pi i)^2} \oint_{|z| = |w| = 1}  \frac{dzdw}{z w} \frac{(1-z^2)(1-w^2)}{(1 - t/(wz))^5(1 - (tw)/z)^5 (1 - (tz)/w)^5 (1 - twz)^5} \\
&  + \frac{1}{2}\frac{1}{2\pi i} \oint_{|z| =  1}  \frac{dz}{z}\frac{(1-z^2)}{(1 - t^2)^5 (1 - t/z)^5 (1 - t z)^5},
\end{align*} 
which we evaluate using the residue theorem, obtaining
\begin{align*}
H(\C[V]^{O(4)}, t) = \frac{1}{2} \left( \frac{1+t^2+6 t^4+t^6+t^8}{\left(1-t^2\right)^{14}} + \frac{1+3 t^2+t^4}{\left(1-t^2\right)^{12}} \right) = \frac{1+t^2+t^4+t^6+t^8}{\left(1-t^2\right)^{14}}.
\end{align*}
We also obtain the Hilbert series for when $G = SO(4,\C)$ for free by only considering the component connected to the identity
\begin{align*}
H(\C[V]^{SO(4)}, t) =  \frac{1+t^2+6 t^4+t^6+t^8}{\left(1-t^2\right)^{14}}.
\end{align*}
To include an arbitrary permutation group $P \subseteq S_n$ acting on the $n$ momenta, one again needs to average over the conjugacy classes of $P$ as discussed previously. 

\section*{Acknowledgements}
We thank Scott Melville and other members of the Cambridge Pheno Working Group for
helpful advice and comments. 
This work has been partially supported by STFC consolidated grants
ST/P000681/1 and 
ST/S505316/1. WH is supported by the Cambridge Trust.

\bibliographystyle{JHEP-2}
\bibliography{generatorsII}

\providecommand{\href}[2]{#2}\begingroup\raggedright\begin{thebibliography}{10}

\bibitem{Weyl}
H.~Weyl, {\em {The classical groups: their invariants and representations}}.
\newblock Princeton University Press, 1966.

\bibitem{LPIPIII}
B.~Gripaios, W.~Haddadin and C.~Lester, {\it Lorentz and permutation invariants
  of particles {III}}, . To appear.

\bibitem{LPIPI}
B.~Gripaios, W.~Haddadin and C.~Lester, {\it {Lorentz and permutation
  invariants of particles {I}}},  \href{http://arXiv.org/abs/2003.05487}{{\tt
  2003.05487}}.

\bibitem{Melia}
B.~Henning, X.~Lu, T.~Melia and H.~Murayama, {\it Operator bases,
  {$S$}-matrices, and their partition functions},  {\em Journal of High Energy
  Physics} (2017).

\bibitem{Hochster}
M.~Hochster and J.~L. Roberts, {\it Rings of invariants of reductive groups
  acting on regular rings are cohen-macaulay},  {\em Advances in Mathematics}
  (1974).

\bibitem{dk}
H.~Derksen and G.~Kemper, {\em {Computational invariant theory}}.
\newblock Springer Berlin Heidelberg, 2002.

\bibitem{Buchberger}
B.~Buchberger, {\it A theoretical basis for the reduction of polynomials to
  canonical forms},  {\em {ACM} {SIGSAM} Bulletin} (1976).

\bibitem{Jacobi1}
R.~Ehrenborg and G.~Rota, {\it Apolarity and canonical forms for homogeneous
  polynomials},  {\em European Journal of Combinatorics} (1993).

\bibitem{Jacobi2}
M.~Beecken, J.~Mittmann and N.~Saxena, {\it Algebraic independence and blackbox
  identity testing},  \href{http://arXiv.org/abs/1102.2789}{{\tt 1102.2789}}.

\bibitem{CoxLittleOShea}
D.~Cox, J.~Little and D.~O'Shea, {\em {Ideals, varieties, and algorithms: an
  introduction to computational algebraic geometry and commutative algebra}}.
\newblock Springer New York, 2008.

\bibitem{M2}
D.~R. Grayson and M.~E. Stillman, {\it \texttt{Macaulay2}, a software system
  for research in algebraic geometry},
  \href{http://arXiv.org/abs/http://www.math.uiuc.edu/Macaulay2/}{{\tt
  http://www.math.uiuc.edu/Macaulay2/}}.

\bibitem{TopPartners}
B.~Gripaios, T.~Müller, M.~Parker and D.~Sutherland, {\it Search strategies
  for top partners in composite higgs models},  {\em JHEP} {\bf 08} (2014) 171.

\end{thebibliography}\endgroup


\providecommand{\href}[2]{#2}\begingroup\raggedright\endgroup

\end{document}